\documentclass[aps,10pt,prl,twocolumn,amssymb,amsmath,amsfonts,showpacs,floatfix,citeautoscript,preprintnumbers,footinbib,groupedaddress,a4paper]{revtex4-1}
\usepackage{times}
\usepackage[latin1]{inputenc}
\usepackage{graphicx}

\usepackage{xspace}

\begin{document}

\title{The distribution of the ratio of consecutive level spacings in random matrix ensembles}
\def\lptmsa{Univ. Paris-Sud, CNRS, LPTMS, UMR8626, 91405 Orsay, France.}
\author{Y.~Y.~Atas}
\author{E.~Bogomolny}
\author{O.~Giraud}
\author{G.~Roux}
\affiliation{\lptmsa}

\date{\today}
\pacs{05.45.Mt, 02.10.Yn, 02.50.-r}

\begin{abstract}
  We derive expressions for the probability distribution of the ratio
  of two consecutive level spacings for the classical ensembles of
  random matrices. This ratio distribution was recently introduced to
  study spectral properties of many-body problems, as, contrary to the
  standard level spacing distributions, it does not depend on the
  local density of states. Our Wigner-like surmises are shown to be
  very accurate when compared to numerics and exact calculations in
  the large matrix size limit. Quantitative improvements are found
  through a polynomial expansion. Examples from a quantum many-body
  lattice model and from zeros of the Riemann zeta function are
  presented.
\end{abstract}

\maketitle

Random matrix theory (RMT) was introduced half a century ago in order
to describe statistical properties of energy levels of complex atomic
nuclei~\cite{porter}. Since then, it has proven to be very useful in a
great variety of different fields \cite{mehta, oxford}.

In quantum chaos~\cite{reviewRMT}, RMT accurately accounts for the
spectral statistics of systems whose classical counterpart is
chaotic. While for quantum Hamiltonians which classical counterpart is
integrable, the Berry-Tabor conjecture~\cite{berry} states that their
level statistics follows a Poisson law, Bohigas, Giannoni and Schmit
conjectured~\cite{bohigas} that the case of quantum Hamiltonians with
chaotic classical dynamics must fall into one of the three classical
ensembles of RMT. These three ensembles correspond to Hermitian random
matrices whose entries are independently distributed respectively real
(GOE), complex (GUE) or quaternionic (GSE) random variables (see
\cite{mehta} for details).

Universality of RMT means that random matrix ensembles describe energy
levels of real systems at a statistical level, and only in a local energy
window when the mean level density is set to unity. Different models
may and do have very different level densities and to compare usual
spectral correlation functions like the nearest-neighbor spacing
distribution one has to perform a transformation called unfolding
\cite{porter,mehta}.  The unfolding procedure consists in changing
variables from the true levels, $e_n$, to new ones, $\overline{e}_n =
\overline{\mathcal{N}}(e_n)$, where $\overline{\mathcal{N}}(e)$ is the
mean number of levels less than $e$, obtained either by smoothing over
many realizations in the case of disordered systems, or by local
smoothing over an energy window large compared to the level spacing
but small compared to variations of $\overline{\mathcal{N}}(e)$. The
unfolded spectrum has automatically a mean level spacing equal to one,
and its statistical properties can thus be directly compared with
those of RMT. When a functional form of $\overline{\mathcal{N}}$ is
known (as for billiards), or when large enough statistics is available,
the unfolding is straightforward and easily implemented.

The situation is different for many-body problems, where
$\overline{\mathcal{N}}(e)$ increases as a stretched exponential
function of energy \cite{bethe} with, in general, unknown lower-order
terms, and where it is difficult to calculate a large number of
realizations because of an exponential increase of the Hilbert space
dimension with the number of particles. In order to circumvent these
difficulties which greatly diminish the precision of statistical tests
in systems with a large number of particles, Oganesyan and Huse
\cite{OgaHus07} proposed a new quantity defined as follows. Let $e_n$
be an ordered set of energy levels and $s_n=e_{n+1}-e_{n}$ the
nearest-neighbor spacings. Oganesyan and Huse considered the
distribution of the ratios $\tilde{r}_n$ defined by
\begin{equation}
\tilde{r}_n=\frac{\mathrm{min}(s_n,s_{n-1})}{\mathrm{max}(s_n,s_{n-1})}=\mathrm{min}\left ( r_n,\frac{1}{r_n} \right )\ ,
\label{r_O_H}
\end{equation}
where 
\begin{equation}
\label{defr}
r_n = \frac{s_n}{s_{n-1}}\ .
\end{equation}
This quantity has the advantage that it requires no unfolding since
ratios of consecutive level spacings are independent of the local
density of states. Such a distribution thus allows a more transparent
comparison with experiments than the traditional level spacing
distribution. For this reason, many recent works use this quantity in
different contexts of many-body systems. As an example let us mention
quantum quenches, where the tools of RMT and quantum chaos were used
as a phenomenological approach to quantify the distance from
integrability on finite size lattices
\cite{Kollath10,Santos10,Collura12}, and also to investigate
numerically many-body localization \cite{OgaHus07,ioffe}.  In these
papers the distribution of consecutive level spacing ratios
$P(\tilde{r})$ was shown to yield more precise results than the usual
spacing distribution $P(s)$.

Although the distribution $P(r)$ plays a more and more
important role in the interpretation of numerical data in quantum
many-body Hamiltonians, only numerical estimates of it exist, and they
are restricted to the GOE ensemble. RMT predictions for $P(r)$
are lacking. Such predictions are essential, both for understanding
its shape for the three RMT ensembles, and for providing accurate
estimates with simple formulas that could be used as an efficient
tool.

This letter fills this gap by providing several important results on
$P(r)$. First, we compute Wigner-like surmises for all three classical
RMT ensembles, which already provide simple analytical formulae in
very good agreement with exact numerics and analytical expressions in
the large matrix size limit. Second, the remaining small differences
are shown to be well fitted to numerical precision by a rather simple
polynomial expansion. Results are then applied to examples on a
quantum many-body Hamiltonian and to zeros of the Riemann zeta
function.


\emph{The ratio of consecutive level spacings distribution} -- Instead
of the quantity \eqref{r_O_H}, we find it more natural to consider
directly the ratio of two consecutive level spacings \eqref{defr} and
its probability distribution $P(r)$. Indeed, let $\rho(e_1,e_2,e_3)$
be the probability density of three consecutive levels with $e_1\leq
e_2\leq e_3$. Assuming translation invariance,
$\rho(e_1,e_2,e_3)=P(s_1,s_2)$ where $s_{i}=e_{i+1}-e_i$. Then
\begin{eqnarray}
P(r)&\equiv &\int P(s_1,s_2)\delta\left (r-\frac{s_1}{s_2}\right )\mathrm{d}s_1\mathrm{d}s_2\nonumber\\
&=&\int_0^{\infty}P(r s_2,s_2)s_2 \mathrm{d}s_2\ .
\label{ps1s2}
\end{eqnarray}
It is physically natural and can be proved analytically that for all
classical RMT ensembles in the bulk of the spectrum (as well as for
Poisson variables) the function $P(s_1,s_2)$ is symmetric, that is,
$P(s_1,s_2)=P(s_2,s_1)$.  This left-right symmetry implies then that
the distributions of $r_n$ and $1/r_n$ are the same, so that $P(r)$
satisfies the following functional equation
\begin{equation}
\label{symetrie}
P(r)=\frac{1}{r^2}P\left(\frac{1}{r}\right)\;.
\end{equation}
Whenever \eqref{symetrie} holds, it is equivalent to consider the
whole distribution $P(r)$ or to restrict the study to the support
$[0,1]$ by considering the variable $\tilde{r}$ defined in
\eqref{r_O_H}, as was done in \cite{OgaHus07}. Here we concentrate on
the whole distribution $P(r)$; since $P(\tilde{r})=2P(r)\Theta(1-r)$,
our results can easily be translated to the restricted
distribution. The integrable (Poisson) case trivially yields
$P(r)=1/(1+r)^2$. We now address the behavior of $P(r)$ for RMT
ensembles.


\emph{Wigner-like surmise} -- For Gaussian ensembles, the joint
probability distribution of $N$ eigenvalues $e_i$ is given
by~\cite{mehta}
\begin{equation}
\label{joint}
\rho(e_1,\ldots, e_N) 
= C_{\beta,N} \!\!\! \prod_{1\leq i<j\leq N}|e_i-e_j|^{\beta}\prod_{i=1}^{N}e^{-\beta e_i^2 / 2},
\end{equation}
where $C_{\beta,N}$ is a known normalization constant and $\beta$ is
the Dyson index equal to 1 (GOE), 2 (GUE) or 4 (GSE). The exact
calculation of $P(r)$ via Eq.~\eqref{ps1s2} requires the calculation
of $P(s_1,s_2)$. Though this calculation is possible from
\eqref{joint} (as shown at the end of this Letter), it ultimately
requires the use of numerical methods and is not transparent. Exactly
the same problem appears in the calculation of the usual
nearest-neighbor spacing distribution, $P(s)$, which is the
probability that the distance between two consecutive levels is
$s$. Rather than cumbersome exact calculations, Wigner derived a
simple approximate expression for $P(s)$,
\begin{equation}
\label{psrmt}
P_W(s)=a_{\beta} s^{\beta} e^{-b_{\beta} s^{2}}\;,
\end{equation}
with some explicitly known normalization constants $a_{\beta}$ and
$b_{\beta}$ \cite{mehta}. This formula, called the Wigner surmise,
corresponds to the exact result for $2\times 2$ matrices, and is in
very good agreement with the exact large-$N$ expressions
\cite{Dietz1990}.

In a similar spirit, we obtain a formula for the ratio distribution 
of two consecutive spacings by performing the exact calculation for 
$3\times 3$ matrices, starting
from the joint distribution \eqref{joint} for three eigenvalues
$e_1,e_2,e_3$. If for instance $e_1\leq e_2\leq e_3$, the ratio $r$ is
given by $(e_3-e_2)/(e_2-e_1)$. Consequently, the distribution $P(r)$
in the $3\times 3$ case is proportional to
\begin{equation*}
\int_{-\infty}^{\infty}\!{d}e_2\int_{-\infty}^{e_2}\!{d} e_1\int_{e_2}^{\infty}\!{d}e_3\ 
\rho(e_1,e_2,e_3)\ \delta\left(r-\frac{e_3-e_2}{e_2-e_1}\right) .
\end{equation*}
After the change of variables $x=e_2-e_1, y=e_3-e_2$, the integration
over $e_2$ is trivial and the remaining integrals read
\begin{equation*}
\iint_{0}^{\infty}\!{d}x{d}y\,\delta(rx-y)x^{\beta+1}y^{\beta}(x+y)^{\beta}e^{-\frac12(x^2+y^2)+\frac16(x-y)^2}.
\end{equation*}
After performing the integrals, the surmise takes the simple form
\begin{equation}
\label{pder}
P_W(r)=\frac{1}{Z_{\beta}}\frac{(r+r^2)^{\beta}}{(1+r+r^2)^{1+\frac{3}{2}\beta}},
\end{equation}
with $Z_{\beta}$ the normalization constant (see values in
Table~\ref{tab:coef}).

\begin{table}[b]
\centering
\begin{tabular}{|c|c|c|c|c|}
\hline
Ens.    & Poisson & GOE      &   GUE & GSE  \\
\hline\hline
$Z_{\beta}$ & - & $\frac{8}{27}$  &  $\frac{4}{81}\frac{\pi}{\sqrt{3}}$ & $\frac{4}{729}\frac{\pi}{\sqrt{3}}$ \\
\hline
$c_{\beta}$ & - & $2\frac{\pi-2}{4-\pi}$  & $4\frac{4-\pi}{3\pi-8}$  &  $8\frac{32-9\pi}{45\pi-128}$ \\
$C$ & - & $0.233378$  & $0.578846$  &  $3.60123$ \\
\hline
$\langle{r}\rangle_W$ & $\infty$ & $ \frac{7}{4}$ & $ \frac{27}{8}\frac{\sqrt{3}}{\pi}-\frac 1 2$ & $ \frac{243}{80}\frac{\sqrt{3}}{\pi} - \frac 1 2$ \\
                      &          & $= 1.75$         & $\approx 1.360735$                      & $\approx 1.174661$ \\
$\langle{r}\rangle_{\text{fit}}$ & - & $1.7781(1)$   & $1.3684(1)$       & $1.1769(1)$   \\
\hline
$\langle{\tilde{r}}\rangle_W$ & $2\ln 2-1$ & $4-2\sqrt{3}$ & $ 2\frac{\sqrt{3}}{\pi}-\frac 1 2$ & $ \frac{32}{15}\frac{\sqrt{3}}{\pi}-\frac 1 2$ \\
                              & $\approx 0.38629$ & $\approx 0.53590$ & $\approx 0.60266$ & $\approx 0.67617$ \\
$\langle{\tilde{r}}\rangle_{\text{fit}}$ & - & $0.5307(1)$ & $0.5996(1)$ & $0.6744(1)$ \\
\hline\hline
\end{tabular}
\caption{Values of useful constants and averages $\langle{r}\rangle$ and $\langle{\tilde{r}}\rangle$. Averages $\langle.\rangle_W$ are calculated from Eq.~\eqref{pder}, and $\langle.\rangle_{\text{fit}}$ from data in Fig.~\ref{pderRMT}.}
\label{tab:coef}
\end{table}   

One can check that this result satisfies the symmetry
\eqref{symetrie}. The distribution $P_W(r)$ has the same level
repulsion at small $r$ than $P(s)$, namely $P_W(r) \sim r^\beta$,
while for large $r$ the asymptotic behavior is 
$P_W(r) \sim r^{-(2+\beta)}$, contrary to the fast
exponential decay of $P(s)$. This surmise also yields an analytic
expression for the mean-values $\langle{r}\rangle_W$ and
$\langle{\tilde{r}}\rangle_W$ widely used in the literature as a
measure of chaoticity (see Table~\ref{tab:coef} for the exact
values).
 
\begin{figure}[t]
\centering
\includegraphics[width=0.82\linewidth,clip]{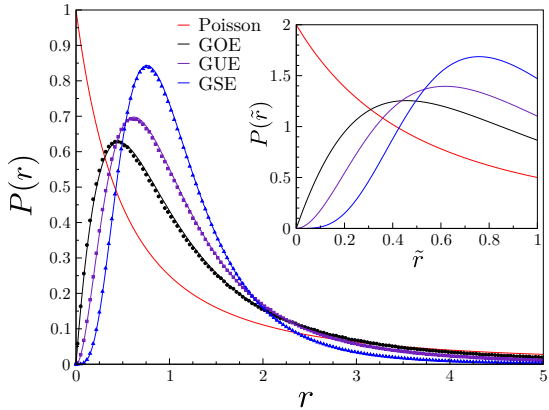}
\caption{(Color online) Distribution of the ratio of consecutive level
  spacings $P(r)$ for Poisson and RMT ensembles: full lines are the
  surmise Eq.~\eqref{pder}, points are numerical results obtained by
  diagonalizing matrices of size $N=1000$ with Gaussian distributed
  entries, averaged over $10^5$ histograms. Inset: the distribution
  $P(\tilde{r})$.}
\label{pderRMT}
\end{figure}

\begin{figure}[t]
\centering
\includegraphics[width=\linewidth,clip]{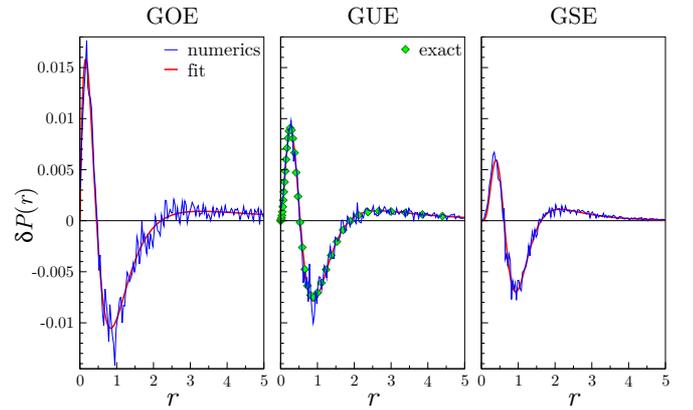}
\caption{(Color online) Difference $\delta P(r) = P_{\text{num}}(r) -
  P_W(r)$ between the numerics and the surmise~\eqref{pder}. The fit
  function is given by Eq.~\eqref{fit}. Green diamonds are results of
  exact calculations obtained from \eqref{paxb} for GUE.}
\label{error}
\end{figure}

\emph{Comparison with numerics and polynomial fit} -- We now
investigate the accuracy of the surmise \eqref{pder} with respect to
numerical calculations for large matrix sizes. As illustrated in
Fig.~\ref{pderRMT}, the surmise is almost indistinguishable from
numerics and can thus be used for practical purposes as a reference to
discriminate between regular and chaotic dynamics. The absolute
difference $\delta P(r) = P_{\text{num}}(r)-P_W(r)$ between numerics
and the surmise \eqref{pder} is plotted in Fig.~\ref{error} for the
three ensembles, and has a maximum relative deviation of about 5\%,
similar to the Wigner surmise for $P(s)$~\cite{Dietz1990}.

In order to go beyond the surmise \eqref{pder}, we propose a simple
expression which perfectly fits this remaining difference $\delta
P(r)$ within our computational accuracy.  In order to fulfill
Eq.~\eqref{symetrie}, and assuming that $P(r)$ for large $N$ and
$P_W(r)$ have the same asymptotic behavior for small and large $r$, a
reasonable ansatz is the following expansion
\begin{equation}
\label{fit}
\delta P_{\text{fit}}(r) = \frac{C}{(1+r)^2}
\left[ \left(r+\frac{1}{r}\right)^{-\beta}\!\!\!\! - c_{\beta} \left(r+\frac{1}{r}\right)^{-(\beta+1)} \right]\ ,
\end{equation}
where $c_{\beta}$ is easily calculated from the normalization
condition $\int_0^{\infty }\delta P(r) dr=0$ (see Table~\ref{tab:coef}
for the exact value). Thus the large-$N$ expression for $P(r)$ can be
fitted by the expression $P(r)=P_W(r)+\delta P_{\text{fit}}(r)$ with
only one fitting parameter, which is the overall magnitude $C$ of the
discrepancy. The best fit $C$ can be found in
Table~\ref{tab:coef}. The corresponding curves are shown in
Fig.~\ref{error}. Thanks to these very good fits, one can quickly
infer accurate predictions for $\langle{r}\rangle$ and
$\langle{\tilde{r}}\rangle$ and any average weighted by $P(r)$ (see
Table~\ref{tab:coef}).\\

\emph{Large-$N$ calculation} -- We now turn to the exact calculation
of $P(r)$ for GUE (i.e. $\beta=2$) in the limit $N\to\infty$,
following a path similar to the derivation of the exact level spacing
distribution $P(s)$.

Our starting point is Eq.~5.4.29 of Ref.~\onlinecite{mehta}. From that
equation, one can check that the probability $p(-t,y,t)$ of having
three consecutive levels at points $-t,y,t$ can be rewritten as
\begin{equation}
\label{paxb}
p(-t,y,t)=\det(1-K)\det[R(x,z)_{x,z=-t,y,t}],
\end{equation}
where $R(x,y)$ is the resolvent kernel, i.~e.~the kernel
of the operator $(1-K)^{-1}K$, and $\det(1-K)$ is the Fredholm
determinant of $K$. Operator $K$ is an integral operator whose action
is defined as
\begin{equation}
\label{oper}
(Kf)(x)=\int_{-t}^{t}K(x,y)f(y)\mathrm{d}y
\end{equation}
with the kernel 
\begin{equation}
K(x,y)=\frac{\sin \pi(x-y)}{\pi(x-y)}\ .
\label{kernel}
\end{equation}
It is known (see e.~g.~\cite{tracy}) that for a kernel of this form
the resolvent kernel can be written as
\begin{equation}
R(x,y)=\frac{Q(x)P(y)-Q(y)P(x)}{x-y},
\end{equation}
with functions $Q(x)$ and $P(x)$ obeying integral equations
\begin{eqnarray}
&&Q(x)-\int_{-t}^{t}K(x,y)Q(y)\mathrm{d}y=\frac{\sin \pi x}{\pi} ,\nonumber\\
&&P(x)-\int_{-t}^{t}K(x,y)P(y)\mathrm{d}y=\cos \pi x .
\label{QP}
\end{eqnarray}
Function $Q(x)$ and $P(x)$ have many useful properties which allow to
relate the calculation of spectral statistics for standard RMT
ensembles to solutions of Painlev\'e equations (see e.~g.~\cite{tracy}
and references therein). Though this approach is elegant, it still
requires numerical resolution of Painlev\'e V equation for $\det(1-K)$
with subsequent solutions of linear equations for $Q(x)$ and $P(x)$
whose coefficients are determined by that solution.

\begin{figure}[t]
\centering
\includegraphics[width=\linewidth,clip]{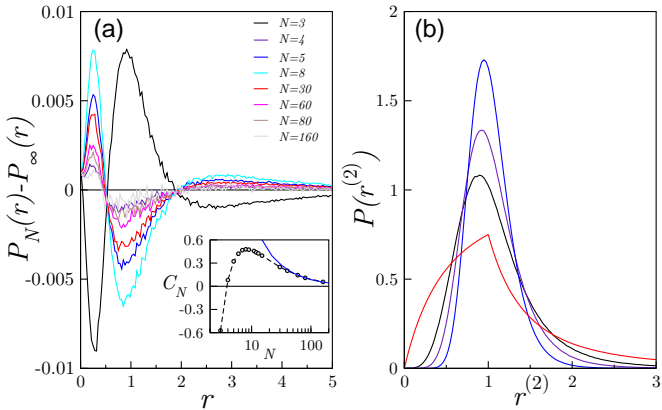}
\caption{(Color online) (a) $P(r)-P_{\infty}(r)$ for GUE and various
  matrix sizes. Inset: constant $C_N$ from the fit \eqref{fit} as a
  function of matrix size $N$ (solid line is a fit $1/N$). (b) Density
  distributions for the overlapping ratio $r^{(2)}_n =
  (e_{n+2}-e_{n})/(e_{n+1}-e_{n-1})$ for Poisson variables and for the
  three classical RMT ensembles (same color code as in
  Fig.~\ref{pderRMT}).
  \label{error2}}
\end{figure}

We find it simpler to use the direct method proposed in
\cite{bornemann} for computing $\det(1-K)$. It is based on a
quadrature method for numerical evaluation of the integrals
\begin{equation}
\label{discretization}
\int_{-t}^{t} f(x)dx =\sum_{k=1}^m w_k f(x_k)
\end{equation}   
appearing in the definition \eqref{oper} of the integral operator
$K$. Such a discretization allows to approximate the determinant of
the integral operator as a finite $m\times m$ determinant
\begin{equation}
\det(1-K)\approx \det \big ( \delta_{jk}-K(x_j,x_k)w_k \big )
\end{equation}
and functions $Q(x)$ and $P(x)$ defined in \eqref{QP} can be obtained
by solving a linear system of $m$ equations. As noted in
\cite{bornemann} the method quickly converges. The result is presented
in Fig.~\ref{error}, where the Clenshaw-Curtis method with up to 60
points of discretization has been used for the discretization
\eqref{discretization}. Figure \ref{error2} (left) shows how the
numerical results converge to the analytic large-$N$ calculation. As
mentioned previously, the fit $P(r)=P_W(r)+\delta P_{\text{fit}}(r)$
works well for all $N$, with an overall $N$-dependent constant $C_N$
in \eqref{fit}. This constant, which gives the amplitude of the
departure from the Wigner-like surmise, asymptotically decreases as
$1/N$ (see inset of Fig.~\ref{error2}).

\begin{figure}[t]
\centering
\includegraphics[width=0.8\linewidth,clip]{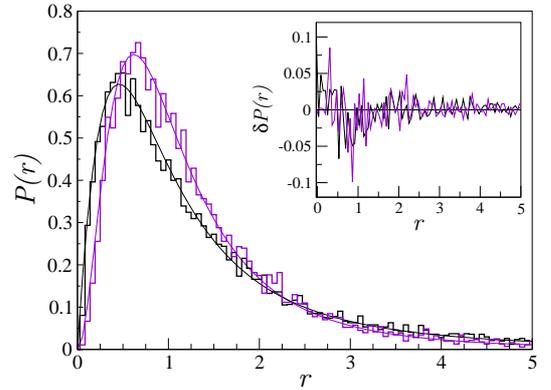}
\caption{(Color online) Histogram of the ratio of consecutive level
  spacings $P(r)$. Black: Quantum Ising model in fields
  $\lambda=\alpha=0.5$ in sector of eigenvectors of $\hat{T}$ with
  eigenvalue $\omega_3$ ($\omega_j=\exp (2\mathrm{i}\pi
  j/L),\hspace*{0.1cm} j=0,1,\ldots ,L-1$), for $L=18$ spins
  (dimension of eigenspace = 14541).  Violet: The same for zeros of
  Riemann zeta function up the critical line ($10^4$ levels starting
  from the $10^{22}$th zero, taken from \cite{odlyzko}). Full lines
  correspond to the Wigner-like surmise Eq.~(\ref{pder}) with
  respectively $\beta=1$ and $\beta=2$. Inset: Difference between the
  numerics and these surmises.}
\label{riemann}
\end{figure}

\emph{Applications} -- To illustrate the above formalism, we
investigate the spectral properties of a quantum Ising chain of $L$
spins$-\frac12$ with periodic boundary conditions in transverse
field $\lambda$ and longitudinal field $\alpha$. The Hamiltonian is
given by
\begin{equation}
\hat{H}=-\sum_{n=1}^L\left(\hat{\sigma}_{n}^{x}\hat{\sigma}_{n+1}^{x} +\lambda\hat{\sigma}_{n}^{z}+\alpha\hat{\sigma}_{n}^{x}\right), \qquad \hat{\sigma}_{L+1}^{x}=\hat{\sigma}_{1}^{x}
\label{non_integrable_model}
\end{equation}
where $\hat{\sigma}_{n}^{x,z}$ are the Pauli matrices at site
$n$. This model recently attracted attention due to its experimental
realization in cobalt niobate ferromagnet \cite{experiment}. The
Hamiltonian (\ref{non_integrable_model}) commutes with the operator
$\hat{T}$ which translates the state by one lattice spacing and obeys
$\hat{T}^L=1$. Consequently, $\hat{H}$ takes a block diagonal form in
the basis of eigenstates of $\hat{T}$, and one has to consider
separately each sector of symmetry. The result for one sector is
illustrated in Fig.~\ref{riemann}. Other symmetry sectors give similar
results. As expected, $P(r)$ agrees well with the GOE prediction
\eqref{pder} with $\beta=1$.

Another example of application is to look at non-trivial zeros of the
Riemann zeta function
\begin{equation}
\zeta(s)=\sum_{n=1}^{\infty}\frac{1}{n^{s}}.
\end{equation}
It is well established that statistical properties of Riemann zeros
are well described by the GUE distribution \cite{riemann}. The
probability distribution of the ratio of two consecutive spacings of
these zeros, presented in Fig.~\ref{riemann}, is in a
perfect agreement with GUE formula \eqref{pder} with $\beta=2$.

\emph{Conclusion} -- The investigation of spectral statistics in
many-body problems with a large number of particles attracted wide
attention in recent years. The absence of a well established
expression for the mean density of states greatly diminishes the
usefulness of standard correlation functions such as the
nearest-neighbor spacing distribution. To avoid this problem, a new
statistical tool has been proposed in \cite{OgaHus07}, namely the
distribution of the ratio of two consecutive level spacings.
 
The main result of the paper is the derivation of simple approximative
formulae for this distribution for classical RMT ensembles. The
resulting Wigner-like surmises agree very well with direct numerical
calculations. The difference between the surmise and the exact
calculations is small and can be fitted by a one-parameter polynomial
formula with excellent accuracy. 

In the same spirit, several different ratios can be introduced which
generalize the quantity \eqref{defr}.  Analytic expressions and
Wigner-like surmises can be derived in a similar way for the density
distributions of these quantities, and will be discussed elsewhere. An
example is given in Fig.~\eqref{error2} (right). All these
distributions are universal in the sense that they apply without any
unfolding or renormalization to spectra ranging from many-body systems
to Riemann zeta function.

YYA was supported by the CFM foundation. GR acknowledges support from
grant ANR-2011-BS04-012-01 QuDec.


\end{document}